\documentclass[cameraready]{Interspeech}

\title{Speaker Identity in Non-Verbal Vocalizations: Conditional Distillation and Mixture of Experts Approach}

\author[affiliation={1}, orcid=0009-0005-9638-310X, equalcontribution]{Tzu-Chieh}{Wei}

\author[affiliation={2}, orcid=0009-0007-3994-6433, equalcontribution]{Yi-Cheng}{Lin}

\author[affiliation={3},orcid=0000-0003-2125-5689]{Huang-Cheng}{Chou}

\author[affiliation={2},orcid=0009-0001-7256-7263]{Kuan-Yu}{Chen}

\author[affiliation={3},orcid=0009-0003-7071-7664]{Hsin-Yen}{Sung}

\author[affiliation={3},orcid=0000-0002-1052-6204]{Shrikanth}{Narayanan}

\author[affiliation={2,4}, orcid=0000-0002-9654-5747, correspondingauthor]{Hung-yi}{Lee}

\address{
    $^1$ University of Michigan, USA, $^2$ National Taiwan University, Taipei, Taiwan\\
    $^3$ Signal Analysis and Interpretation Laboratory (SAIL), University of Southern California, USA \\
    $^4$ National Taiwan University Artificial Intelligence Center of Research Excellence, Taipei, Taiwan 
}

\email{jeff20020302@gmail.com, even.dlion8@gmail.com, tlkagkb93901106@gmail.com}


\usepackage{comment}
\usepackage{booktabs}
\usepackage{multirow}
\usepackage{graphicx}
\usepackage{subcaption}

\usepackage{pgfplots}
\usepackage{subcaption}
\usepgfplotslibrary{groupplots}
\usetikzlibrary{calc}
\pgfplotsset{compat=1.18}

\begin{document}

\maketitle
\begin{abstract}
As expressive text-to-speech (TTS) and voice conversion (VC) systems increasingly generate non-verbal vocalizations (NVVs) to enhance naturalness, reliable speaker verification (SV) becomes essential to objectively assess identity consistency across both verbal and non-verbal segments. Yet current SV systems generalize poorly to NVVs, and fine-tuning on NVV data causes catastrophic forgetting of speech performance. We present the first systematic study across 10 NVV types and propose a framework combining frozen Data2Vec self-supervised features with ECAPA-TDNN, enhanced by a Mixture of Experts (MoE) module with learned domain-aware routing. A conditional distillation loss on speech inputs via a pretrained teacher retains speech-to-speech accuracy, while a contrastive loss bridges the speech-NVV domain gap. Our method reduces speech-NVV EER from 38.93\% to 22.66\% over a pretrained baseline, and improves speech EER from 13.17\% to 9.24\% via distillation.\footnote{Codebase: https://github.com/wiizzz/nonverbal-sv}
\end{abstract}

\section{Introduction}
Modern Text-to-Speech (TTS)~\cite{ju2024naturalspeech, du2025cosyvoice, hsu2025breezyvoiceadaptingttstaiwanese, shi2026emotion} and Voice Conversion (VC)~\cite{liu2024zero} systems achieve near-human naturalness for modal speech, but generating realistic non-verbal vocalizations (NVVs) such as laughter, coughing, and breathing remains an open challenge.  
Since these generated sounds must maintain speaker identity, objective evaluation demands automated, scalable  Speaker Verification (SV) capable of assessing identity consistency across both verbal and non-verbal segments, a requirement that human perceptual studies cannot feasibly meet at scale.

Currently, the dominant SV paradigm leverages Self-Supervised Learning (SSL) front-ends (e.g., WavLM~\cite{chen2022wavlm}, HuBERT~\cite{hsu2021hubert}) coupled with discriminative backends like ECAPA-TDNN~\cite{desplanques2020ecapa}, achieving strong performance on standard corpora like VoxCeleb~\cite{Nagrani_2017_Voxceleb1}. 
Extending to a full NVV taxonomy poses a harder multi-domain problem: unlike speech, where shared phonemic structure provides a common representational anchor, different NVV categories such as coughing~\cite{zhang2017speaker}, laughter~\cite{lin2023laugh} exhibit distinct acoustic signatures \cite{sauter2010perceptual} with no unifying structural processing framework. Moreover, these non-phonemic inputs violate the implicit design prior of current SV systems, which assume that all inputs contain phonemically structured speech. An SV system must therefore accommodate multiple heterogeneous acoustic domains, each diverging considerably from modal speech in spectral and temporal properties. Whether modern SV representations can generalize under such acoustic diversity remains unexplored.

As we demonstrate empirically in Section~\ref{subsec:pilot}, existing SV systems exhibit substantially increased overlap between same-speaker and different-speaker score distributions when evaluated on NVV pairs. 
While naively fine-tuning these models on non-verbal datasets might intuitively bridge this domain gap, it introduces a critical side effect: severe degradation of normal speech-to-speech verification performance, an issue largely overlooked in previous NVV literature.

To address this without sacrificing baseline speech performance, we propose a framework combining Mixture of Experts (MoE) routing with conditional knowledge distillation. 
The acoustic heterogeneity across NVV categories motivates an MoE design~\cite{shazeer2017outrageously}: rather than forcing a single feature pathway to handle both phonemically structured speech and diverse non-phonemic vocalizations, we insert MoE modules into the standard SSL + ECAPA-TDNN pipeline, with dedicated expert subnetworks for speech and non-verbal inputs, respectively. This binary routing enables each expert to specialize in their respective acoustic domain. Furthermore, drawing inspiration from knowledge distillation used in continual learning~\cite{li2017learning}, we propose a novel conditional \textbf{distillation loss}. 
This mechanism explicitly forces the model to retain the speech verification capabilities of the pretrained baseline while learning non-verbal representations.

Our main contributions are summarized as follows:
\begin{itemize}
    \item To the best of our knowledge, this is the first systematic evaluation of modern SV systems across a diverse taxonomy of 10 NVV types, going beyond prior work that examined only isolated event categories
    \item We propose a novel conditional distillation loss that effectively utilizes pretrained SV models to learn non-verbal representations while substantially mitigating standard speech capabilities.
    \item We design an MoE-based SV system that separates non-verbal and speech processing paths, providing a speaker similarity evaluation tool for next-generation expressive TTS and VC systems.
\end{itemize}

\begin{figure}[t] 
    \centering
    \includegraphics[width=\linewidth]{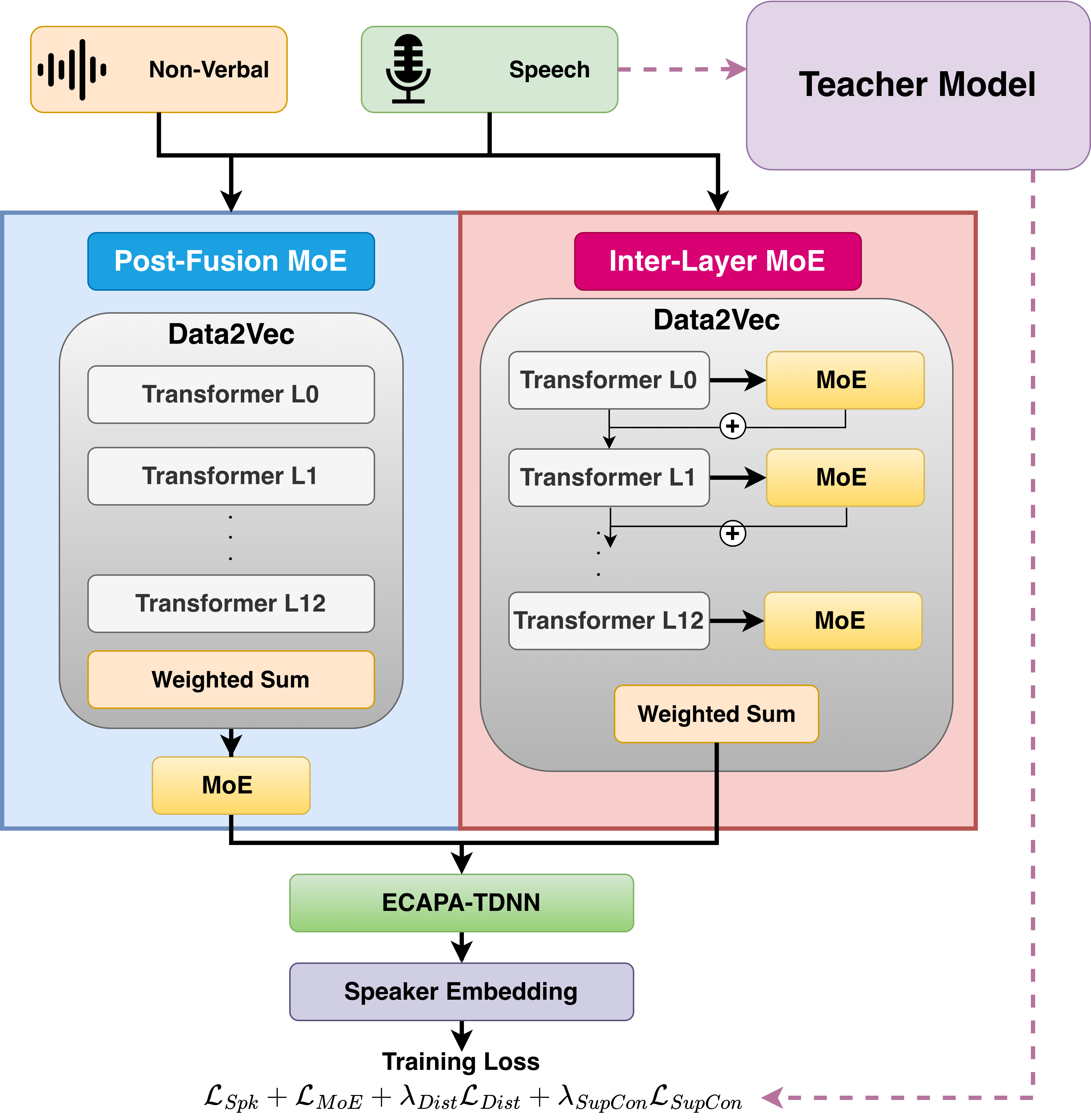}
    \caption{Overall pipeline of proposed method.}
    \label{fig:non_verbal}
    \vspace{-8mm}
\end{figure}


\section{Methodology}
\vspace{-1mm}

As illustrated in Figure~\ref{fig:non_verbal}, our proposed framework comprises three core components: (1) a frozen Data2Vec and ECAPA-TDNN backbone; (2) an MoE module for domain-aware feature routing; and (3) a multi-objective training strategy integrating conditional distillation, contrastive bridging, and domain-aware routing constraints.

\vspace{-1mm}
\subsection{Backbone Architecture: Data2vec + ECAPA-TDNN}
\label{ssec:backbone}
\vspace{-1mm}
The core architecture comprises an SSL feature extractor and a speaker embedding backend. For the front-end, we adopt Data2vec~\cite{Baevski_2022_data2vec}, a modality-agnostic model predicting continuous latent representations via self-distillation. This avoids the phonemic discretization bottleneck of cluster-based approaches~\cite{chen2022wavlm, hsu2021hubert}, a choice empirically validated by its superior NVV verification performance in Table~\ref{tab:main_results}. The extracted frame-level representations are then fed into an ECAPA-TDNN~\cite{desplanques2020ecapa} backend. Its multi-scale aggregation captures both local and global characteristics, effectively modeling the diverse temporal structures of NVVs, from brief coughs to extended laughter.

\subsection{Feature Integration: Mixture of Experts (MoE)}
\label{ssec:moe}

To separate representational pathways for speech and NVVs, we introduce an MoE module. Given a frame-level representation $\mathbf{h} \in \mathbb{R}^d$, a gating network conditionally routes it to specialized experts via $\mathbf{g} = \mathrm{TopK}(\mathrm{softmax}(\mathbf{W}_g \mathbf{h}), k)$. The output is the weighted sum $\mathbf{y} = \sum_{i \in \mathrm{TopK}} g_i \cdot E_i(\mathbf{h})$, with $k=2$ in all experiments.

We evaluate two integration strategies (Figure~\ref{fig:non_verbal}) to bridge the frozen SSL and ECAPA-TDNN models. \textbf{Design 1: Post-Fusion MoE} applies late-stage adaptation via a single MoE layer operating on the weighted sum of the frozen SSL hidden states. Conversely, \textbf{Design 2: Inter-Layer Residual MoE (IR-MoE)} inserts trainable MoE adapters after each frozen transformer block, propagating adapted representations forward.

\subsection{Training Objectives}
\label{ssec:training_objective}
Our joint optimization framework integrates standard SV pipeline, event-guided MoE routing constraints, and conditional knowledge distillation to ensure domain-aware specialization and mitigate catastrophic forgetting. The total loss is defined as:
\begin{equation}
    \footnotesize
    \mathcal{L} = \mathcal{L}_{Spk} + \mathcal{L}_{MoE} + \lambda_{Dist} \mathcal{L}_{Dist} + \lambda_{SupCon} \mathcal{L}_{SupCon},
\end{equation}
where $\mathcal{L}_{Spk}$ is the standard AAM-softmax loss~\cite{deng2019arcface} and $\mathcal{L}_{MoE}$ encapsulates our proposed routing constraints, formulated as:
\begin{equation}
    \footnotesize
    \mathcal{L}_{MoE} = \lambda_{Bal}\mathcal{L}_{Bal} + \lambda_{Intra}\mathcal{L}_{Intra} + \lambda_{Inter}\mathcal{L}_{Inter}.
\end{equation}

\vspace{2mm}
\noindent\textbf{Event-Guided MoE Routing Constraints ($\mathcal{L}_{MoE}$):}
To enforce expert specialization and prevent routing collapse, we apply three distinct penalties. First, we maximize the entropy of the mean routing distribution using a batch-level load balancing loss~\cite{fedus2022switch}. Second, we apply event-aware soft regularization using train-time labels. By maintaining an Exponential Moving Average (EMA) routing prototype, $\mu_e$, for each acoustic event $e$~\cite{grill2020bootstrap}, we enforce intra-event consistency via Kullback-Leibler (KL) divergence:
\begin{equation}
    \mathcal{L}_{Intra} = \mathbb{E}_{(x,e)}[\mathrm{KL}(\mathbf{P(x)} \parallel \mu_e)],
\end{equation}
where $\mathbf{P(x)}$ denotes the routing distribution for input $\mathbf{x}$. Finally, we ensure inter-event separation via a cosine-margin loss:
\begin{equation}
    \mathcal{L}_{Inter} = \sum_{e \neq e'} \max(0, \text{sim}(\mu_e, \mu_{e'}) - \delta),
\end{equation} 
where $\text{sim}(\cdot, \cdot)$ denotes cosine similarity, and $\delta$ is a predefined margin. Together, these constraints implicitly guide the router to distinguish varying non-verbal events without hard-coding expert assignments.

\vspace{2mm}
\noindent\textbf{Conditional Knowledge Distillation ($\mathcal{L}_{Dist}$):}
To selectively preserve speech representations without forcing NVVs into an ill-suited acoustic manifold, we leverage a frozen WavLM-based SV model~\cite{chen2022wavlm} as a teacher. We conditionally apply this distillation loss based on utterance type. Let $\mathbf{e}^i_s$ and $\mathbf{e}^i_t$ denote the student and teacher embeddings for the $i$-th sample in a batch. The loss is formulated as the cosine distance:
\begin{equation}
    \mathcal{L}_{Dist} = \frac{1}{|S|}\sum_{i \in S} \left(1-\frac{\mathbf{e}^i_s \cdot \mathbf{e}^i_t}{||\mathbf{e}^i_s|| \cdot ||\mathbf{e}^i_t||}\right),
\end{equation}
where $S$ denotes the index set of speech samples within the batch. This formulation ensures the model retains robust structural knowledge of modal speech while allowing unconstrained adaptation for non-verbal representations. During inference, utterance-type labels are not required; the MoE gating network, optimized via event-guided constraints, autonomously routes speech and NVV inputs to their respective specialized experts.

\vspace{2mm}
\noindent\textbf{Supervised Contrastive Domain Bridging ($\mathcal{L}_{SupCon}$):}
To explicitly bridge the acoustic domain gap between speech and NVV, we incorporate a supervised contrastive loss~\cite{khosla2020supervised, dixit2024improving}. The loss is defined as:
\begin{equation}
    \footnotesize
    \mathcal{L}_{\mathrm{SupCon}} = \sum_{i=1}^{N} \frac{-1}{\lvert \mathcal{P}(i) \rvert} \sum_{p \in \mathcal{P}(i)} \log \frac{\exp\ ( \mathrm{sim}(z_i, z_p) / \tau )}{\sum_{a\in\mathcal{A}(i)} \exp\ ( \mathrm{sim}(z_i, z_a) / \tau)},
\end{equation}
where N is the number of samples in the batch and $\tau$ is a temperature scalar. For an anchor $z_i$, $\mathcal{P}(i)$ and $\mathcal{A}(i)$ represent the index sets of positive and negative samples within the batch, respectively. Crucially, to enforce a shared speaker manifold across acoustic domains, positive pairs ($\mathcal{P}(i)$) include not only intra-domain samples but also cross-domain speech-NVV pairs from the same speaker. Negatives are sampled from different speakers.

\begin{table*}[t]
    \centering
    \fontsize{8}{10}\selectfont 
    \caption{\small Overall performance comparison. NvS, NvN, SvS. All fine-tuned baselines and proposed models share identical setups and base objectives (AAM-Softmax and Contrastive Loss). The proposed models additionally incorporate conditional distillation.}
    \vspace{-3mm}
    \label{tab:main_results}
    \begin{tabular}{llcccccc}
    \toprule
    \multirow{2}{*}{\textbf{Setting}} & \multirow{2}{*}{\textbf{Model Architecture}} & \multicolumn{2}{c}{\textbf{NvS}} & \multicolumn{2}{c}{\textbf{NvN}} & \multicolumn{2}{c}{\textbf{SvS}} \\
    \cmidrule(lr){3-4} \cmidrule(lr){5-6} \cmidrule(lr){7-8}
     &  & \textbf{EER (\%)} & \textbf{minDCF} & \textbf{EER (\%)} & \textbf{minDCF} & \textbf{EER (\%)} & \textbf{minDCF} \\
    \midrule
    Zero-shot & wavlm-base-plus-sv & 38.93 & 0.998 & 39.13 & 0.997 & \textbf{5.60} & \textbf{0.352} \\
    \midrule
    \multirow{6}{*}{Self-trained} 
     & ECAPA-TDNN (Fbank) & 28.14 & 0.937 & 33.57 & 0.953 & 11.48 & 0.665 \\
     & WavLM + ECAPA-TDNN & 27.58 & 0.917 & 28.63 & 0.916 & 10.92 & 0.571 \\
     & Voc2Vec + ECAPA-TDNN & 26.16 & 0.881 & 31.34 & 0.881 & 11.54 & 0.605 \\
     & Data2Vec + ECAPA-TDNN & \underline{23.33} & \textbf{0.806} & \underline{27.98} & \underline{0.834} & 10.76 & 0.611 \\
     & (Voc2Vec+WavLM) + ECAPA & 26.7 & 0.880 & 29.78 & 0.877 & 11.22 & 0.550 \\
     & (Voc2Vec+Data2Vec) + ECAPA & 23.47 & 0.834 & 28.56 & \textbf{0.818} & 10.36 & 0.570 \\
     & (Data2Vec+WavLM) + ECAPA & 25.46 & 0.848 & 28.40 & 0.839 & 9.24 & 0.528 \\
    \midrule
    \multirow{2}{*}{Proposed} 
     & MoE-1 (Data2Vec, 4 experts) & 23.95 & 0.833 & 28.38 & 0.857 & \underline{9.00} & 0.527 \\
     & \textbf{MoE-2 (Data2Vec, 4 experts)} & \textbf{22.66} & \underline{0.817} & \textbf{27.52} & 0.867 & 9.24 & \underline{0.525} \\
    \bottomrule
    \end{tabular}
    \vspace{-6mm}
\end{table*}

\section{Experiment}



\begin{table}[!t]
\caption{Distribution of NVV types in the NonverbalTTS dataset. The dataset is heavily imbalanced, with \textit{Breath} comprising over 67\% of all NVV samples.}
\label{tab:nvv_dist}
\vspace{-3mm}
\centering
\fontsize{8}{10}\selectfont 
\begin{tabular}{lccccc}
\toprule
\textbf{Type} & Breath & Laugh & Cough & Sniff & Sigh \\
\textbf{\#Samples}   & 3,102   & 953   & 194   & 186   & 140  \\
\midrule
\textbf{Type} & Groan & Throat & Snore & Sneeze & Grunt \\
\textbf{\#Samples}   & 111   & 104    & 11    & 9      & 7    \\
\bottomrule
\end{tabular}
\vspace{-6mm}
\end{table}

\begin{table}[!b]
    \vspace{-4mm}
    \centering
    \fontsize{8}{10}\selectfont 
    \caption{\small Ablation of different distillation losses on IR-MoE.}
    \vspace{-3mm}
    \label{tab:loss_ablation}
    \begin{tabular}{cccc}
    \toprule
    {\textbf{Loss Components}} & \textbf{NvS} & \textbf{NvN} & \textbf{SvS} \\
    \cmidrule(lr){1-2}
     \textbf{Distill}& \textbf{EER (\%)} & \textbf{EER (\%)} & \textbf{EER (\%)} \\
    \midrule
     $\times$ & 24.95 & 29.61 & 13.17 \\
    \checkmark & \textbf{22.66} & \textbf{27.52} & \textbf{9.24} \\
    \bottomrule
    \end{tabular}
\end{table}

\subsection{Database and Pre-processing}
To systematically investigate the domain mismatch in SV models, we utilize the NonverbalTTS~\cite{borisov2025nonverbaltts} dataset for our experimental evaluations. It comprises 17 hours of audio split into training (1,314 speakers), validation (46 speakers), and test (147 speakers) partitions, with no speaker overlap across splits. The annotations span 10 NVV categories, with the detailed sample distribution across these classes provided in Table \ref{tab:nvv_dist}. For trial construction, each testing segment is evaluated against all available positive (same-speaker) utterances and 30 randomly selected negative (different-speaker) utterances. This results in 18,043 trials for NVV vs. Speech (NvS), 18,398 for NVV vs. NVV (NvN), and 10,764 for Speech vs.Speech (SvS).

To prepare the data for training our specialized nonverbal SV model, we used the Montreal Forced Aligner (MFA) ~\cite{mcauliffe2017montreal} to obtain word‑level alignments, and segmented each utterance into isolated NVV clips and speech segments. This preprocessing yields the separated speech and NVV data streams required by both the MoE routing and the conditional distillation framework.

\subsection{Baseline Models}
We compare our proposed model against seven baselines that all share the ECAPA‑TDNN backend. These include one system using conventional Fbank features, three single‑SSL front‑ends (WavLM, Data2Vec, and Voc2Vec~\cite{Koudounas_2025_voc2vec}), and three dual‑SSL fusion configurations (WavLM + Voc2Vec, Data2Vec+Voc2Vec, and WavLM + Data2Vec), as summarized in Table \ref{tab:main_results}.

\begin{figure}[h]
    \centering
    \includegraphics[width=\linewidth]{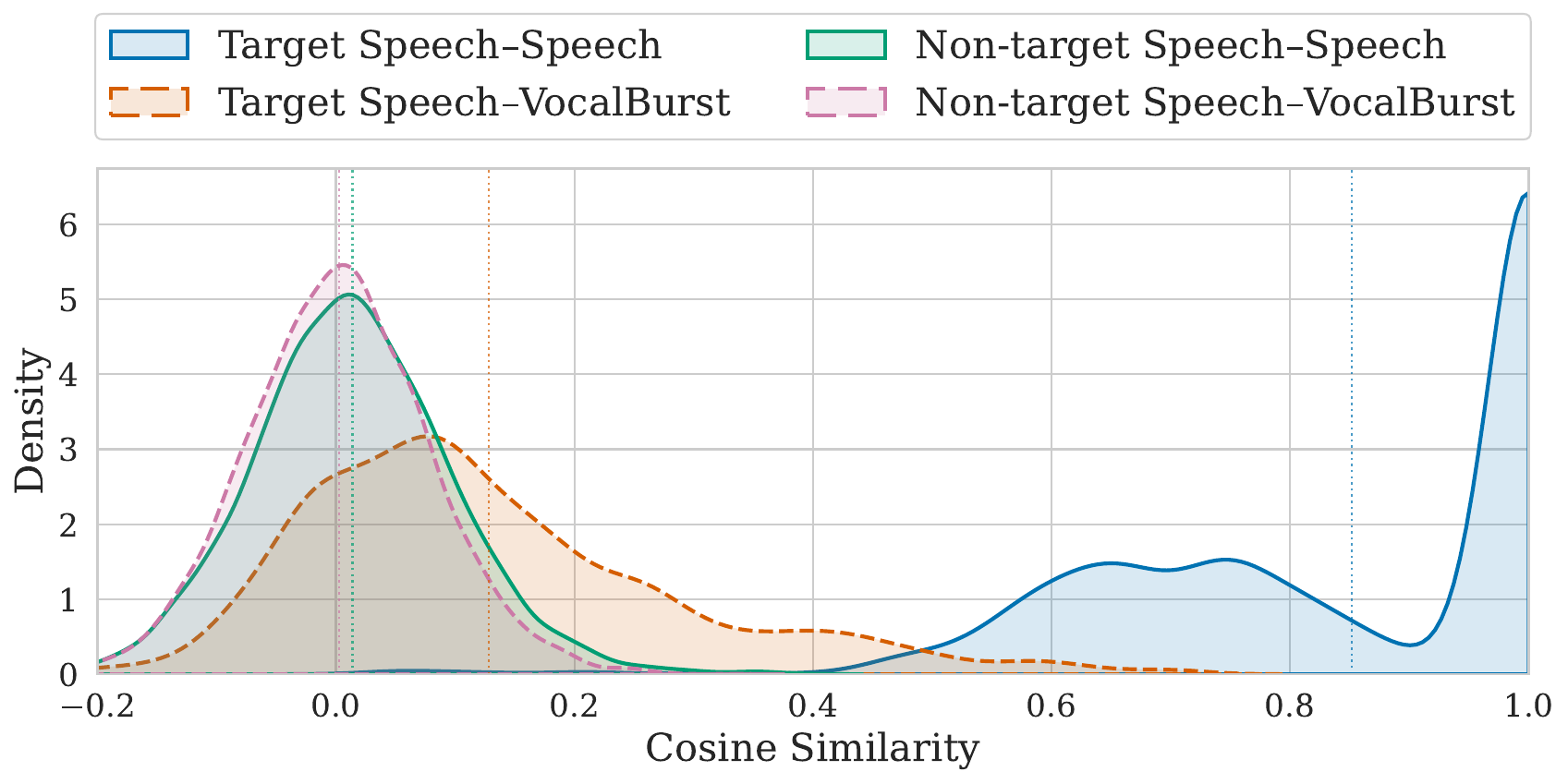}
    \caption{\small Cosine similarity distributions illustrating the domain gap between speech and vocal bursts. Target speech–vocal burst trials (dashed orange) exhibit substantial overlap with non-target distributions, indicating an increased risk of false rejections in ECAPA-TDNN.}
    \label{cos_sim_exp}
    \vspace{-6mm}
\end{figure}

\vspace{-1mm}
\subsection{Training Details} 
\vspace{-1mm}
The model parameters are optimized using the Adam optimizer with a weight decay of \(1\times10^{-4}\). We adopt a cosine annealing learning rate schedule, initializing the learning rate at \(5\times10^{-3}\) and gradually decaying it to \(1\times10^{-4}\) to facilitate stable convergence. Training is supervised by the AAM-Softmax loss, with the margin and scale set to 0.2 and 32, respectively. For all experiments, the downstream ECAPA-TDNN architecture is fixed with 1024 channels in the convolutional frame-level layers, producing 192-dimensional speaker embeddings. The loss weights are set to \(\lambda_{\text{Dist}}=10\), \(\lambda_{\text{SupCon}}=0.3\), \(\lambda_{\text{Bal}}=0.1\), \(\lambda_{\text{Intra}}=0.05\), and \(\lambda_{\text{Inter}}=0.05\). The temperature parameter \(\tau\) is set to 0.07, the margin threshold \(\delta\) to 0.5, and the batch size to 128.\\
\textbf{Batch Composition Strategy:} To ensure sufficient cross-domain positive pairs for $L_{SupCon}$, we adopt a speaker-balanced batch sampling strategy with a total batch size of 128. Specifically, each batch is composed of 16 unique speakers, where each speaker contributes exactly 8 utterances (a mix of speech and NVV utterances). Furthermore, during the loss computation, our implementation explicitly prioritizes cross-domain positive pairs (same speaker, different modality) and falls back to intra-domain positive pairs (same speaker, same modality) if cross-domain pairs are unavailable for a given anchor.\\
\textbf{Progressive Training Schedule:} 
To stabilize the multi-objective optimization required specifically for the Mixture-of-Experts models, we employ a progressive schedule. We begin with a warm-up phase optimizing solely $\mathcal{L}_{Spk}$, followed by the introduction of MoE routing with a high $\lambda_{Bal}$ to prevent early collapse. Finally, we gradually scale up the event-regularization ($\lambda_{Intra}$, $\lambda_{Inter}$).

\vspace{-1mm}
\subsection{Evaluation Protocol and Metrics}
\vspace{-1mm}
Cosine similarity is used for trial scoring. Verification performances are measured by Equal Error Rate (EER) and the minimum normalized detection cost function (mDCF) with $P_{target} = 0.05$.

\vspace{-1mm}
\section{Results and Analyses}
\vspace{-1mm}
This section systematically evaluates the performance of standard SV models alongside our proposed framework. Specifically, we first quantify the domain mismatch that degrades zero-shot verification performance on non-verbal vocalizations. We then analyze the transfer learning dilemma and demonstrate how our conditional distillation loss mitigates catastrophic forgetting of modal speech. Finally, we validate the effectiveness of the MoE architecture in overcoming representational bottlenecks by dynamically routing verbal and non-verbal features.


\begin{figure}[t]
    \centering
    \begin{tikzpicture}
    \begin{groupplot}[
        group style={
            group size=2 by 1,
            horizontal sep=0.6cm,
            ylabels at=edge left,
        },
        width=0.6\columnwidth,
        height=3.3cm,
        ylabel={EER (\%)},
        xtick={0,1,2,3,4},
        xticklabels={0,1,2,4,8},
        grid=major,
        grid style={dashed, gray!30},
        every axis plot/.append style={thick},
        every mark/.append style={scale=1.2},
        label style={font=\small},
        tick label style={font=\footnotesize},
        title style={font=\small\bfseries, at={(0.5,1.15)}, anchor=north},
    ]
    \nextgroupplot[title={NvS}, ymin=22.0, ymax=24.5, xlabel={}]
    \addplot[color=blue!70!black, mark=*] coordinates {
        (0,23.33) (1,23.45) (2,23.95) (3,22.66) (4,23.05)
    };
    \addplot[only marks, mark=*, mark size=3.5pt, color=red!80!black] coordinates {(3,22.66)};
    \node[font=\scriptsize\bfseries, color=red!80!black, anchor=south west] at (axis cs:3.1,22.66) {22.66};

    \nextgroupplot[title={SvS}, ymin=8.0, ymax=11.5, xlabel={}]
    \addplot[color=blue!70!black, mark=square*] coordinates {
        (0,10.76) (1,10.08) (2,9.00) (3,9.24) (4,8.96)
    };
    \addplot[only marks, mark=square*, mark size=3.5pt, color=red!80!black] coordinates {(4,8.96)};
    \node[font=\scriptsize\bfseries, color=red!80!black, anchor=south west] at (axis cs:4.1,8.96) {8.96};

    \end{groupplot}
    
    \node[font=\small, anchor=north] at ($(group c1r1.south)!0.5!(group c2r1.south) + (0,-0.4cm)$) {Number of IR-MoE Experts};
    \end{tikzpicture}
    \vspace{-3mm}
    \caption{Effect of IR-MoE expert count on verification performance. The 4-expert configuration achieves the best NvS EER, while SvS EER continues to decrease with more experts.}
    \label{fig:moe_experts}
    \vspace{-6mm}
\end{figure}
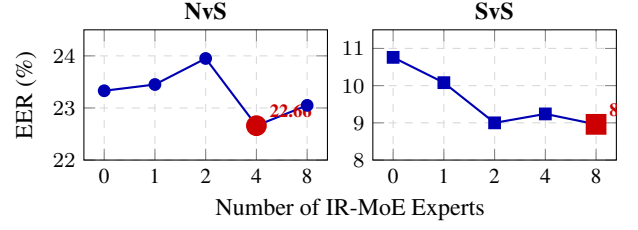

\vspace{-1mm}
\subsection{Domain Mismatch and Transfer Dilemma}
\label{subsec:pilot}
\vspace{-1mm}
To quantify the domain mismatch between modal speech and NVV, we construct 20,000 evaluation trials on the NonverbalTTS test split. These trials are uniformly distributed across four conditions, with exactly 5,000 pairs each: target (same-speaker) and non-target (different-speaker) pairs for both the intra-domain Speech-to-Speech (SvS) and cross-domain NVV-to-Speech (NvS) verification tasks. Utterances shorter than 0.25 seconds are excluded to ensure stable feature extraction. We extract speaker embeddings using a representative pre-trained model, ECAPA-TDNN, and analyze the domain mismatch via cosine similarity distributions (Figure~\ref{cos_sim_exp}).

As shown in the ``Zero-shot'' section of Table~\ref{tab:main_results}, the state-of-the-art \textit{wavlm-base-plus-sv} model~\cite{chen2022wavlm} achieves a strong EER of 5.60\% on the standard SvS task. 
However, performance degrades substantially on NvS and NvN tasks, yielding EERs of 38.93\% and 39.13\%, respectively. 
This confirms that representations learned exclusively from modal speech fail to generalize to the irregular acoustic structures of NVVs.

\vspace{-1mm}
\subsection{Ablation of Objective Functions}
\vspace{-1mm}
\label{ss:ablation_results}
To mitigate the catastrophic forgetting observed during NVV adaptation, we introduce a conditional distillation loss. Table~\ref{tab:loss_ablation} presents an ablation study of the objective functions using our proposed IR-MoE architecture. When optimized solely with AAM-Softmax and supervised contrastive loss, the model struggles to balance both domains, resulting in a degraded SvS EER of 13.17\%. 

Incorporating the distillation loss explicitly constrains the model's speech embeddings to align with the pretrained teacher's representations. This regularization successfully bridges the domain gap, reducing the NvS EER to 22.66\% while recovering standard speech performance (SvS EER improves from 13.17\% to 9.24\%). We used the paired bootstrap test~\cite{Confidence_Intervals} to estimate the 95\% confidence intervals for the EER reductions. The strictly positive intervals across all conditions 
demonstrate that our method yields a statistically significant improvement. These results demonstrate that conditional distillation is a crucial mechanism for maintaining multi-domain robustness. 

Although our approach significantly enhances NvS verification, all fine-tuned configurations, including ours, exhibit a higher SvS EER compared to the zero-shot \textit{wavlm-base-plus-sv} baseline (5.60\%). We attribute this gap to the vast difference in training data scale between VoxCeleb2 and NonverbalTTS. Addressing this discrepancy through large-scale joint training remains a direction for future work.

\vspace{-2mm}
\subsection{Evaluation of MoE Architecture}
\vspace{-1mm}
Standard SSL models process vastly different acoustic statistics through a single feature-extraction pipeline, often leading to representational bottlenecks. As shown in Table~\ref{tab:main_results}, the Data2Vec+ECAPA-TDNN model emerges as the strongest standalone baseline with an NvS EER of 23.33\%, yet it lacks dynamic feature routing.

To address this, our MoE architecture explicitly separates verbal and non-verbal processing paths. We conducted an ablation study on the number of expert networks to determine the optimal routing capacity (Figure~\ref{fig:moe_experts}). 
Expanding the architecture from a single unified path to multiple experts enhances the reliability of the system. Specifically, the 4-expert configuration (denoted as IR-MoE) achieves the best overall balance. 
It minimizes the NvS EER to 22.66\% while maintaining a competitive SvS EER of 9.24\%. Beyond 4 experts, NvS performance slightly degrades (23.05\%), likely because the limited training data (17 hours) provides insufficient samples per expert for effective specialization. Interestingly, SvS EER continues to improve (8.96\%), suggesting that increased capacity better isolates distilled speech representations from non-verbal interference. This demonstrates the effectiveness of decoupled expert routing in multi-domain speaker verification.

\vspace{-2mm}
\section{Conclusion and Future Work}
\vspace{-1mm}
This paper addresses a critical blind spot in current speech research by presenting the first systematic study of speaker identity verification across 10 distinct NVV types. We demonstrate that standard SV models suffer severe acoustic mismatch and catastrophic forgetting when adapted to NVVs. To overcome this, we proposed an Inter-Layer Residual MoE architecture paired with a novel conditional distillation strategy. Experimental results confirm that our framework successfully bridges the Speech-NVV domain gap (reducing EER from 38.93\% to 22.66\%) while substantially mitigating degradation of standard speech verification performance. 
Crucially, our framework employs a divide-and-conquer strategy: decoupling distinct acoustic domains in the intermediate MoE layers while unifying the speaker identity manifold in the final embedding space via $\mathcal{L}_{SupCon}$. Future work will theoretically analyze the representational dynamics of this intermediate-separation and final-fusion mechanism, and explore zero-shot adaptation to evaluate synthetic NVVs directly from modern TTS pipelines.

\section{Acknowledgments}
We acknowledge the National Center for High-Performance Computing (NCHC) of the National
Institutes of Applied Research (NIAR) in Taiwan for providing computing resources. Additionally, this work was supported by the Ministry of Education (MOE) of Taiwan under
the project Taiwan Centers of Excellence in
Artificial Intelligence, through the NTU Artificial
Intelligence Center of Research Excellence (NTU
AI-CoRE), NSTC, Taiwan (Grant No.114-2917-I-564-030 to Huang-Cheng Chou), US NSF (IIS 2311676), and ODNI IARPA ARTS Program (Contract D2023-2308110001).


\section{Generative AI Use Disclosure}

Generative AI tools assisted in polishing the manuscript's language. 
The authors remain solely responsible for the research design, experiments, analysis, and reported results. 
AI tools did not contribute to the substantive scientific content.

\bibliographystyle{IEEEtran}
\bibliography{mybib}

\end{document}